\documentclass[floatfix,showpacs,preprintnumbers,amsmath,amssymb,aps,twocolumn,superscriptaddress,prl,10pt]{revtex4-1}

\usepackage{amsfonts}
\usepackage[pdftex]{graphicx}
\usepackage{color}
\usepackage{bm}
\usepackage{multirow}
\usepackage{comment} 

\newcommand{\figref}[1]{Fig.~\ref{#1}}
\newcommand{\e}[1]{\text{e}^{#1}}

\newcommand{\diffd}{\text{d}}

\renewcommand{\vec}[1]{\mathbf{#1}}

\newcommand{\ket}[1]{|{#1}\rangle}

\newcommand{\ant}[0]{\hat{\mathcal{A}}}

\begin{document}

\title{Ferromagnetic spin correlations in a few-fermion system}
\author{P.O.~Bugnion}
\affiliation{Cavendish Laboratory, J.J. Thomson
  Avenue, Cambridge, CB3 0HE, United Kingdom}
\author{G.J.~Conduit}
\affiliation{Cavendish Laboratory, J.J. Thomson
  Avenue, Cambridge, CB3 0HE, United Kingdom}
\date{\today}

\begin{abstract}
  We study the spin correlations of a few fermions in a quasi
  one-dimensional trap. Exact diagonalization
  calculations demonstrate that repulsive interactions between
  the two species drives ferromagnetic correlations. The
  ejection probability of an atom provides an experimental probe of the
  spin correlations. With more than five atoms trapped, the system
  approaches the itinerant Stoner limit. Losses to Feshbach molecules are
  suppressed by the discretization of energy levels when fewer than seven
  atoms are trapped.
\end{abstract}


\maketitle

Recent experimental advances allow investigators to confine up to twenty
atoms in a single trap and address their quantum
state~\cite{Cheinet08,Serwane11}. This precise control enabled the
Heidelberg group to confirm the fundamental physics of short range
repulsion~\cite{Busch98,Mora04,Idziaszek06,Liu10,Rontani12,
  Brouzos12,Gharashi12,Rubeni12,Rotureau13,Sala13,Bolda05,Melezhik09}. Here
we demonstrate that just such repulsive interactions acting between a few
fermions allows us to construct a Hamiltonian analogous to the Stoner
model~\cite{Stoner38} and offers experimentalists an opportunity to observe
the emergent ferromagnetic correlations without losses to Feshbach
molecules.

The itinerant ferromagnet predicted by the Stoner Hamiltonian has never been
cleanly realized and studied in the solid state.  However, it was
proposed~\cite{Duine05,Conduit08,LeBlanc09} that a ferromagnetic phase could
be created with a fermionic cold atom gas. An experiment by
the MIT group displayed signatures compatible with
ferromagnetism~\cite{Jo09,Conduit09i,Conduit09ii}, but the observations were
later explained by a loss mechanism~\cite{Haller10,Pekker11,Sanner12}. To
circumvent losses it has been suggested that magnetic correlations could be
explored in systems with a mass imbalance~\cite{vonKeyserlingk11},
two-dimensional geometry~\cite{Conduit2D}, spin spirals~\cite{Conduit10}, or
flux lattices~\cite{Baur12}. Here we demonstrate how a quasi one-dimensional
system containing only a few fermions could avoid the problems associated
with losses and display ferromagnetic correlations. Our main result is shown
in \figref{fig:ExpSetup}: the discretization of energy levels in the few
fermion system means that losses to Feshbach molecules are restricted to a
narrow range of interaction strengths, allowing a tunneling measurement of
the ejection of an atom to expose the underlying ferromagnetic correlations.

In this paper we first describe the proposed experimental setup, then
demonstrate the emergence of ferromagnetic correlations that we study
through a tunneling process.  Finally, we show that the formation of
the competing dimer state is inhibited by the discretization of the energy
levels in the harmonic confining potential.

\section{Experimental setup}

\begin{figure}
 \includegraphics[width=0.35\linewidth]{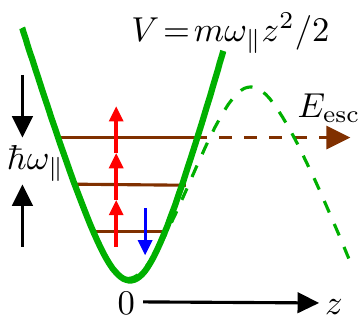}
 \includegraphics[width=0.51\linewidth]{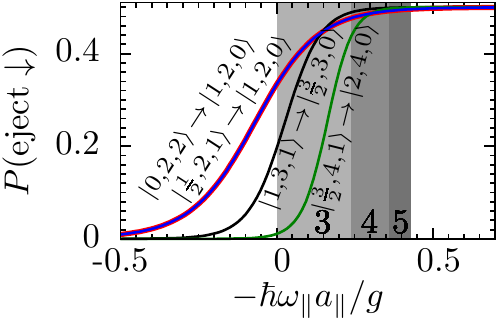}
 \caption{(Color online)
   {\it Left}: The green solid line is the trapping
   potential $V(z)$ that is lowered to the green dashed line to allow an
   atom to escape. The energy levels for up-spin (red) and down-spin (blue)
   atoms within the potential are shown in brown, with the putative escape
   of the highest energy up-spin atom.
   {\it Right}: The ejection probability of a down-spin atom into a state
   fully polarized along the quantization axis. The gray region is excluded
   due to atom losses for differing initial numbers of atoms.}
 \label{fig:ExpSetup}
\end{figure}

A fermionic gas of two hyperfine states with pseudospin
$\sigma\in\{\uparrow,\downarrow\}$ is tightly confined as shown in
\figref{fig:ExpSetup}.  We seek to solve the Hamiltonian $\hat{H}=\sum_i
[-\hbar^{2}\nabla_i^{2}/2m+m\omega_{\perp}^{2}(x_i^{2}+y_i^{2})/2+
m\omega_{\parallel}^{2}z_i^{2}/2 ] +\sum_{i<j}V(\vec{r}_{i}-\vec{r}_{j})$, with
$m$ the atomic mass and $\vec{r}_i=(x_i,y_i,z_i)$ the position of atom $i$. The
confining potential is axially symmetric with trap frequencies
$\omega_{\perp}=10\omega_{\parallel}$~\cite{Serwane11,Zurn12}, and we define
the harmonic oscillator lengths $a_{i}=\sqrt{2\hbar/m\omega_{i}}$.  Only a
single transverse mode is occupied, constraining the atoms into the quasi-1D
regime. We therefore re-parameterize the interspecies potential
$V(\vec{r})=-U\Theta(R-|\vec{r}|)$ first into the s-wave scattering length
$a_{3\text{D}}=R[1-\tan(\chi)/\chi]$ with $\chi=R\sqrt{mU}/\hbar$, and then as
a one-dimensional pseudopotential~\cite{Olshanii98}
$g=\hbar^{2}a_{3\text{D}}/ma_{\perp}(a_{\perp}-Ca_{3\text{D}})$ with
$C=\zeta(1/2)\approx1.46$. A confinement induced resonance emerges at
$a_{3\text{D}}=a_{\perp}/C$. We verified that the results tend to the contact
limit below $R=0.2a_\parallel$. In the limit $\omega_\parallel \to 0$, we
recover the Stoner Hamiltonian $\hat{H}=\sum_i [-\hbar^{2}/2m
    (\partial^2/\partial z_i^2)+
m\omega_{\parallel}^{2}z_i^{2}/2]+\sum_{i<j}g\delta(z_{i}-z_{j})$.

To probe the quantum state we apply a magnetic field gradient to tilt the
external potential (see \figref{fig:ExpSetup}) and allow one atom to
escape. 
We denote the number of trapped spins
$N_{\uparrow}$ and $N_{\downarrow}$. Investigators can directly
measure the spin in the quantization direction
$S_{\text{z}}=(N_{\uparrow}-N_{\downarrow})/2$ and the total number of atoms
$N_{\uparrow}+N_{\downarrow}$ in the final state using the single atom
addressability~\cite{Serwane11}. However, our measure of
ferromagnetic correlations, the spin
$\vec{S}=\langle\sum_{n}(c_{n\uparrow}^{\dagger}~c_{n\downarrow}^{\dagger})
\!\cdot\!\boldsymbol{\sigma}\!\cdot\!
(c_{n\uparrow}~c_{n\downarrow})^{\text{T}}\rangle$ is SU(2) invariant, 
where $\boldsymbol{\sigma}$ denotes the vector of Pauli-spin
matrices and $c_{n\sigma}$ is the annihilation operator of an atom of spin
$\sigma$ from harmonic oscillator state $n$. Therefore,
the spin quantum number defined through
$s(s+1)=\langle\vec{S}^2\rangle$ is a good quantum number, allowing us to
define the quantum state $\ket{s,N_\uparrow,N_\downarrow}$. With two atoms a
polarized $s=1$ state can be generated not only from the $S_{\text{z}}=1$
state, denoted $\ket{1,2,0}$, but also from the $S_{\text{z}}=0$ state
denoted $\ket{1,1,1}$, corresponding to the prototypal triplet states
$\ket{\uparrow\uparrow}$ and
$(\ket{\uparrow\downarrow}+\ket{\downarrow\uparrow})/\sqrt{2}$.  The
unpolarized $S_{\text{z}}=0$ state, denoted $\ket{0,1,1}$, corresponds to
the singlet state
$(\ket{\uparrow\downarrow}-\ket{\downarrow\uparrow})/\sqrt{2}$.

\section{Energy of states}

\begin{figure}[!t]
 \newlength{\stdgap}
 \setlength{\stdgap}{-5.5pt}
 \begin{tabular}{l}
 \includegraphics[width=0.96\linewidth]{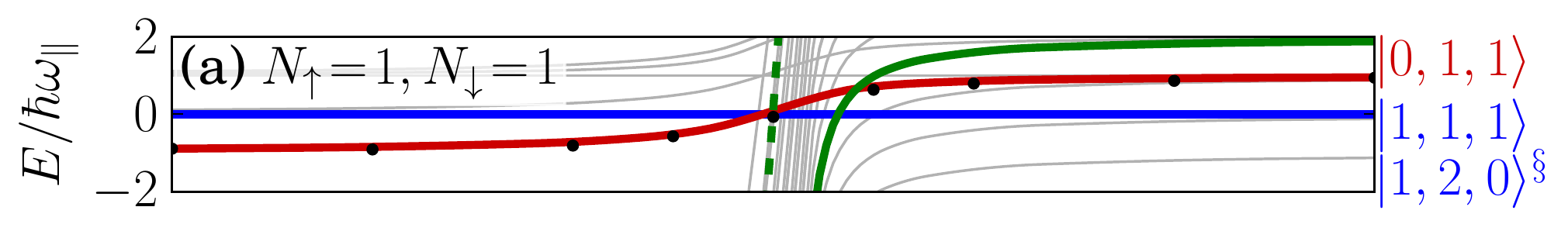}\\[\stdgap]
 \includegraphics[width=0.995\linewidth]{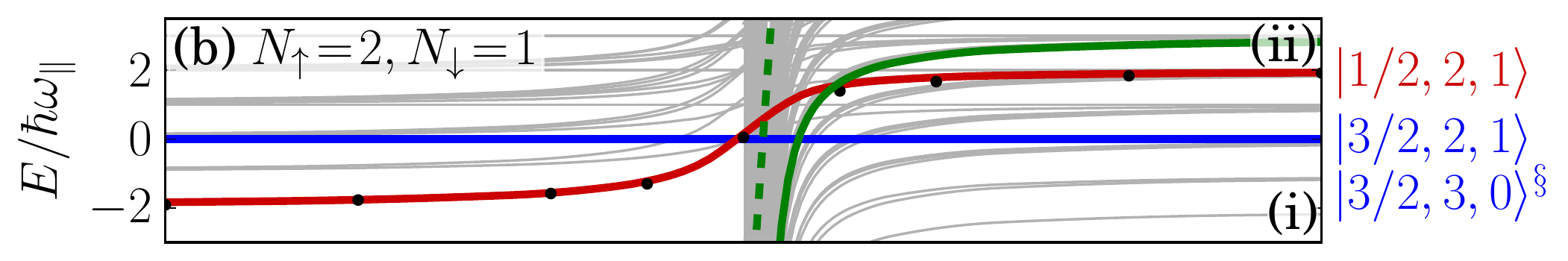}\\[\stdgap]
 \includegraphics[width=0.96\linewidth]{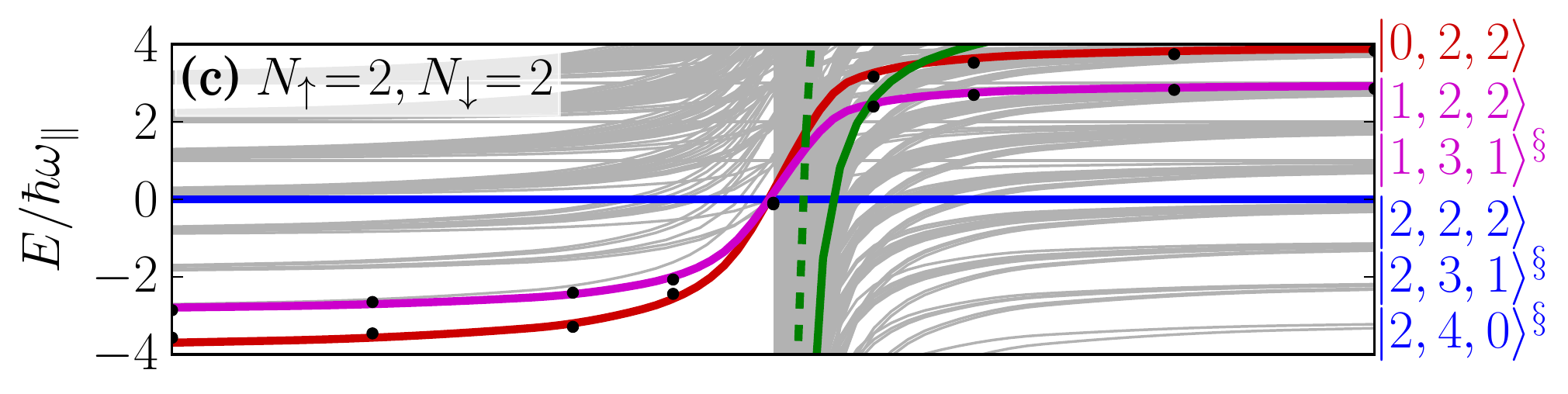}\\[\stdgap]
 \includegraphics[width=0.99\linewidth]{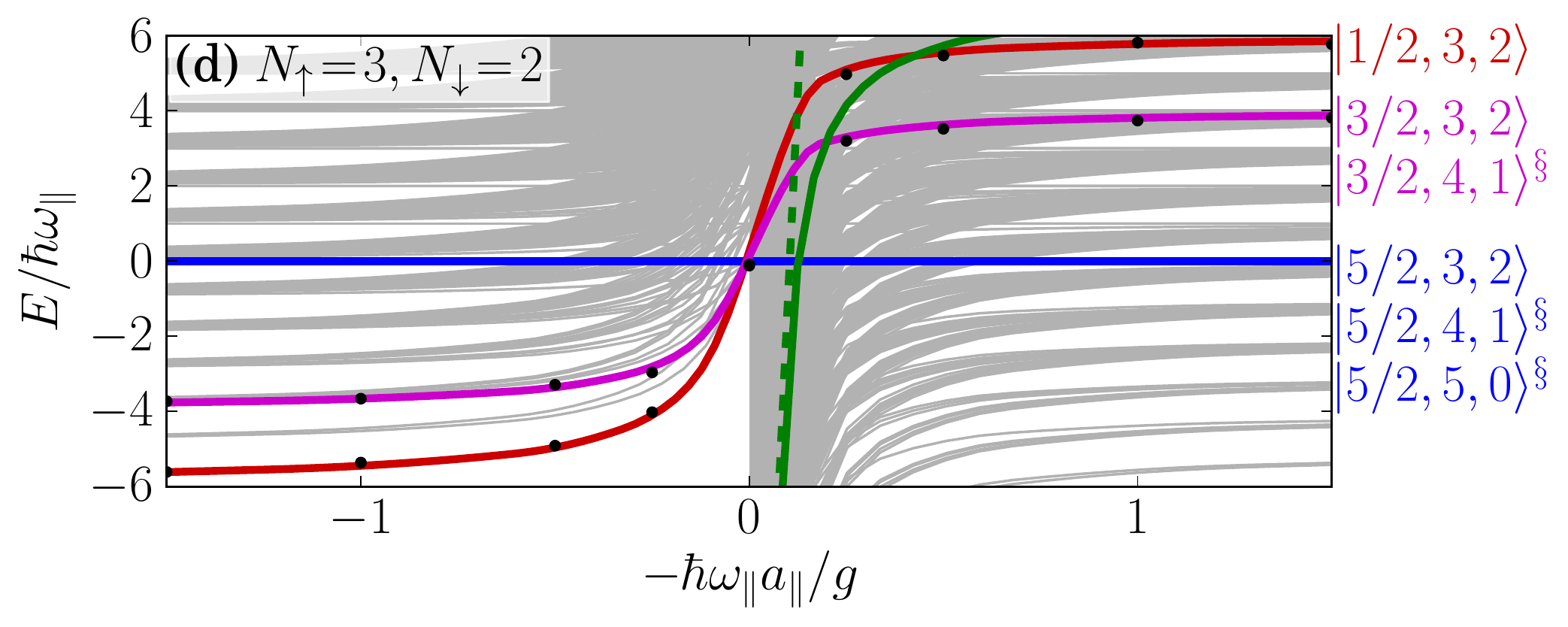}\\[\stdgap]
 \begin{tabular}{lll}
  (e) Distribution&(f) System energy&(g) Pair correlations\\
  \includegraphics[width=0.31\linewidth]{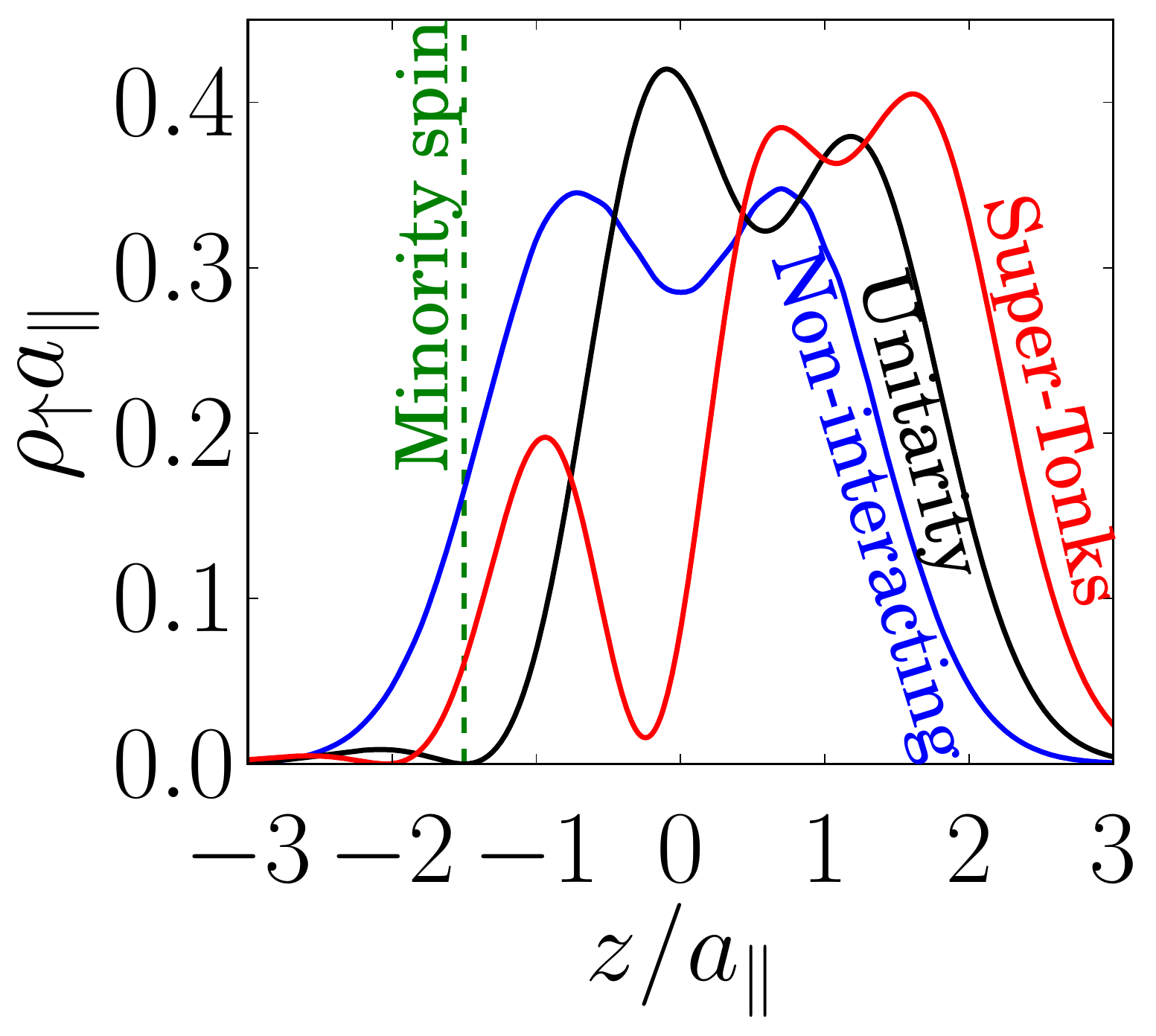}&
  \includegraphics[width=0.31\linewidth]{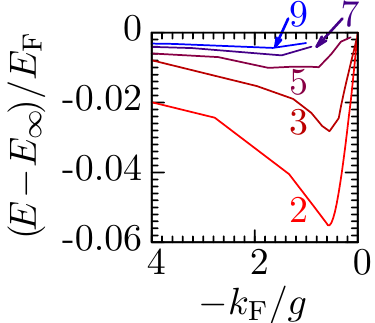}&
  \includegraphics[width=0.31\linewidth]{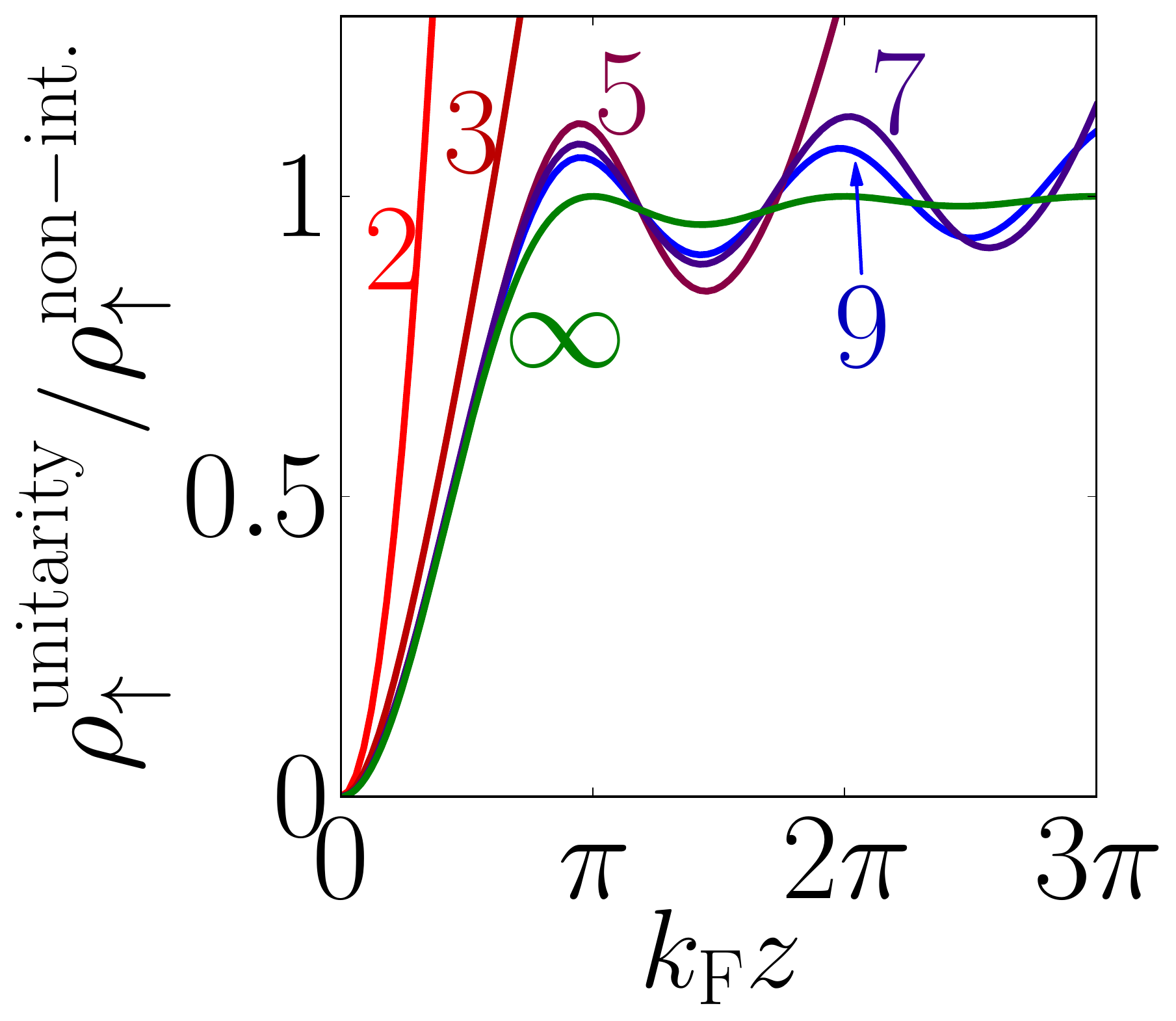}
 \end{tabular}
 \end{tabular}
 \caption{(Color online)
   (a-d) The energy of the (two-five)-atom
   states calculated with exact diagonalization. The red, magenta,
   and blue lines highlight open channels with $\S$
   signifying identical states with larger $S_{\text{z}}$, and gray
   lines indicating all other states. The green line shows the final
   bound molecular state that crosses the low-spin open channel, and
   the green dashed line the bound state with the COM
   motion excited into the $n_{\text{x},\text{y}}=2$ orbital examined
   in Ref.~\cite{Sala13}. The points show QMC results whose
   uncertainty is the point size.
   (e) Density profile of the state $\ket{1/2,2,1}$ for two majority
   species atoms with a pinned minority atom.
   (f) Difference in energy from the polaron in an infinite
   system~\cite{McGuire65}
   with number of trapped atoms.
   (g) Pair correlation function at the confinement induced resonance around
   a down-spin pinned at $z=0$ normalized by the non-interacting density
   profile.}
 \label{fig:TwoParticle}
\end{figure}

Our main tool to study the system is exact diagonalization. We build a
one-atom basis from the Gaussian orbitals
$\phi_{n_\text{x},n_\text{y},n_\text{z}}(x,y,z)$ of the harmonic trapping
potential. We retain all orbitals that satisfy $(n_\text{x},n_\text{y})\le2$
and $n_\text{z}\le20$. We construct the Slater determinants
and select the $10,\!000$ determinants with lowest non-interacting energy,
guaranteeing convergence of the energy of the open channel within
$0.005\hbar\omega_\parallel$.  This is
much smaller than the energy scale of magnetization,
$N_{\uparrow}N_{\downarrow}\hbar\omega_{\parallel}$ and less than the energy
difference from the true itinerant state shown in
\figref{fig:TwoParticle}(f).  We next calculate the interaction matrix
elements numerically, construct the Hamiltonian matrix for a particular $g$
using the Slater-Condon rules, and diagonalize it to obtain the eigenstates
$\{\psi_m(g)\}$.

Exact diagonalization is restricted to systems of fewer than five atoms.
However, the experimental setup can contain up to twenty atoms so to analyze
the general many-body case we use the QMC code
\textsc{casino}~\cite{Needs08,Needs10}.  The approach is a refinement of
that used in previous studies of 
ferromagnetism~\cite{Ceperley80,Ortiz99,Zong02,Conduit09ii,Pilati10,Chang11}.
We use a trial wave function $\psi=FD$ that is a product of a Jastrow factor
$F$ and a Slater determinant, $D=\ant\{\prod_{i\in
  n_\uparrow}\phi_{n_\mathrm{z}}(\vec{r}_i)\} \ant \{\prod_{i\in
  n_\downarrow}\phi_{n_\mathrm{z}}(\vec{r}_i)\}$, where $\ant$ is the
anti-symmetrization operator.  The orbitals are chosen to give the correct non-interacting state on either
side of the confinement induced resonance. The Slater determinant accounts
for fermion statistics while the Jastrow factor includes further
interparticle correlations. To study the open channel and avoid occupation
of the bound state we use the lowest-order constrained variational
method~\cite{Pandharipande73,Pandharipande77} that is common in nuclear
physics and has also been used to study cold atom gases
~\cite{Cowell02,Chang04,Pilati10,Chang11}. This method solves
the Hamiltonian $[-\diffd^2/\diffd r^2+mV(r)]rf(r)=k^2rf(r)$.  For low
energy s-wave scattering this gives $f(r)\approx1-a_{3\text{D}}/r$ that has
a node at the scattering length $a_{3\text{D}}$, and saturates at large
distances. To guarantee occupation of the upper branch, this solution is embedded into a Jastrow factor
$F=\prod_{i,j}f(|\vec{r}_{i}-\vec{r}_{j}|)$ where $\vec{r}_{i}$ is the
position of the $i$th up-spin and $\vec{r}_{j}$ the $j$th down-spin 
~\cite{Pilati10,Chang11}. In the
quasi one-dimensional setting the transverse confinement does not allow
occupation of the bound state in the range of interaction strengths of
interest $0.24\lesssim-\hbar\omega_\parallel a_\parallel/g<\infty$. The
maximum binding energy of the last band to cross the open channel is
$\sim-1.5\hbar\omega_{\parallel}$, which is much less than the energy scale
of the transverse modes $\sim20\hbar\omega_{\parallel}$.  This was
further confirmed by studying the exact diagonalization states, where deep
in the Super-Tonks regime occupation of the higher transverse modes is
$\sim10^{-7}$, resulting in the strong agreement between exact
diagonalization results and QMC demonstrated in
\figref{fig:TwoParticle}(a-d).  Meanwhile, to calculate the binding energy
of the molecule at $g>0$ and the ground open channel state at $g<0$ we use a
Jastrow factor $F=\e{J}$, where $J$ includes the polynomial expansion in
atom-atom separation proposed in Ref.~\cite{Drummond04} with eight
variational parameters.

With the exact diagonalization and QMC formalism in place, in
\figref{fig:TwoParticle}(a-d) we compare the ground state energy predicted
by both exact diagonalization and QMC, and also the two atom exact
analytical solution~\cite{Busch98,Idziaszek06,Rontani12}. There is strong
agreement at all interaction strengths.  The underlying attractive potential
means that exact diagonalization also delivers the multitude of molecular
bound states and repeated bands incremented by $\hbar\omega_{\parallel}$
corresponding to center-of-mass (COM) motion. In the two atom system
\figref{fig:TwoParticle}(a), in the non-interacting
regime, $-\hbar\omega_{\parallel}a_{\parallel}/g\to-\infty$, the lower spin
state $\ket{0,1,1}$ has the lowest energy. At the confinement induced
resonance the spin states cross~\cite{Serwane11,Zurn12}, and in the
Super-Tonks regime, $-\hbar\omega_{\parallel}a_{\parallel}/g\to\infty$, the
$s=1$ states have lower energy.

In the three-body system in \figref{fig:TwoParticle}(b) three open channel
states are possible: the low spin $\ket{1/2,2,1}$ and the high spin states
$\ket{3/2,3,0}$ and $\ket{3/2,2,1}$. Similarly to the two-body system, at
weak interactions the $s=1/2$ state has the lower energy, the bands cross at
the confinement induced resonance, and in the Super-Tonks regime the $s=3/2$
states are favorable, in good agreement with existing
literature~\cite{Brouzos12,Gharashi12}.  In \figref{fig:TwoParticle}(c) we
also studied the four atom case where three values for the spin are
available: $s\in\{0,1,2\}$. The three bands cross at the confinement induced
resonance meaning that any potential onset of ferromagnetism would be
abrupt, as occurs in the infinite body case~\cite{Wang11,Cui12}. With five
atoms the molecular bands become more prevalent, and we will later
demonstrate how they allow losses into bound molecules.

In the Super-Tonks regime the high spin state is energetically favorable so
the gas would enter the magnetic phase if it were not blocked by spin
conservation. However, the spatial distribution
of the atoms betrays the underlying magnetic correlations. 
In the three atom state $\ket{1/2,2,1}$, we pin the minority down-spin atom
at $z=-1.5a_{\parallel}$. In the non-interacting
case the up-spin atom density is concentrated around the trap center
irrespective of the down-spin position. At the confinement induced resonance
the up-spin density is driven to zero at the down spin pinning position,
whereas in the Super-Tonks regime, the up-spin atoms are forced away from the
down-spin forming a separate magnetic domain. 

Now that we have observed the emergence of magnetic correlations we study the
energy of a single down-spin in a trap with $N_{\uparrow}$ majority spin atoms
to assess the consequences of system size and whether the system serves as a
model for the Stoner Hamiltonian. We compare our system to the analytically
solvable polaron limit~\cite{McGuire65} of a single down-spin in a sea of
up-spin atoms. We first study the energy of a polaron in
\figref{fig:TwoParticle}(f). With $N_{\uparrow}=1$, exact diagonalization
displays less repulsion energy than the infinite body case. On inserting more
majority spin atoms the energy quickly tends to the infinite sized
limit~\cite{McGuire65}, being within $\lesssim1\%$ at all interaction strengths
with $N_{\text{tot}}\ge5$. Secondly, in \figref{fig:TwoParticle}(g) we study
the pair correlation function. With more majority spin atoms the pair
correlation function quickly tends to the infinite body limit~\cite{McGuire65},
with the correct correlation hole and first Friedel oscillation observed for
$N_{\text{tot}}\ge5$. Both pieces of evidence indicate that systems with
$N_{\text{tot}}\ge5$ are faithful representations of the
itinerant Stoner Hamiltonian.

\section{Tunneling statistics}

Although the magnetic phase is energetically favorable in the Super-Tonks
regime its formation is prohibited by spin conservation. In
\figref{fig:ExpSetup} we therefore tilt the trap to allow one atom
to escape. This allows the system to tunnel into the magnetic ground state
containing one fewer atom.  We calculate the tunneling rate using Fermi's
Golden rule.  The tunneling rate $\Gamma$ exhibits an exponential dependence on
the escape energy, so we need only consider tunneling from the highest
occupied orbital with maximal energy $E_\text{esc}$. We now consider a general
intermediate interaction strength and calculate the probability of forming a
particular state $i$, $p_{i}=\Gamma_{i}/\sum_{j}\Gamma_{j}$. This tunneling
probability calculated from the exact diagonalization data is shown in
\figref{fig:ExpSetup}. We focus on the polaron limit with multiple up-spin
atoms and a single down-spin atom.

At zero interactions the highest energy majority spin
atom is expelled leading to zero probability of ejecting the minority
spin atom, whereas at the confinement induced resonance all atoms have an
equal probability of expulsion. In the Super-Tonks regime starting with
$N_{\text{tot}}$ atoms, the system will tunnel most rapidly into the state
with lowest energy -- the fully polarized state with
$s=(N_{\text{tot}}-1)/2$.  There are two quantum states available: with
$S_{\text{z}}=(N_{\text{tot}}-1)/2$ formed by the ejection of the down-spin
atom and $S_{\text{z}}=(N_{\text{tot}}-2)/2$ formed by the ejection of an
up-spin atom. Starting with three atoms tunneling into a two atom state,
these are simply the triplet states $\ket{\uparrow\uparrow}$ and
$(\ket{\uparrow\downarrow}+\ket{\downarrow\uparrow})/\sqrt{2}$.  These two
possibilities occur with equal probability, giving a plateau
probability of $1/2$ for ejecting a minority spin atom in the Super-Tonks
regime irrespective of the initial number of atoms.  The probability curves
in \figref{fig:ExpSetup} become increasingly sharp with more atoms
because of the larger energy exchange over the same range of
interaction strengths.

The tunneling method is sensitive to $S_{\text{z}}$ but not $s$ so does not
provide a full diagnosis of the final quantum state. This is exemplified
when starting from the polaron state in the Super-Tonks regime where the
ejection probability of a minority spin is $1/2$ rather than unity. To
distinguish the $\ket{1,1,1}$ state from the other possible $S_{\text{z}}=0$
state, $\ket{0,1,1}$, one could ramp the interaction strength into the
non-interacting regime and measure the energy through a second
tunneling measurement~\cite{Serwane11,Zurn12}. Should the
$\ket{1,1,1}$ state be dominant the energy will be independent of
interaction strength, whereas if $\ket{0,1,1}$ dominates, the energy will
fall.

\section{Loss mechanism}

\begin{figure}
 \begin{tabular}{lll}
 (a) $N_{\!\uparrow}\!\!=\!8$, $N_{\!\downarrow}\!\!=\!1$&
 (b) Crossing map&
 (c) Critical $g$\\[\stdgap]
 \includegraphics[width=0.31\linewidth]{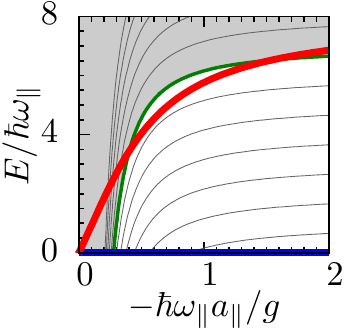}&
 \includegraphics[width=0.31\linewidth]{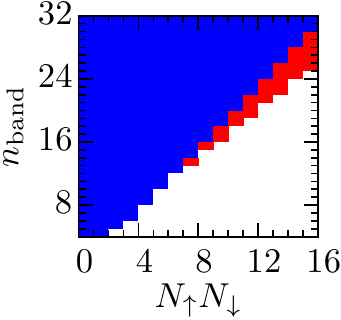}&
 \includegraphics[width=0.31\linewidth]{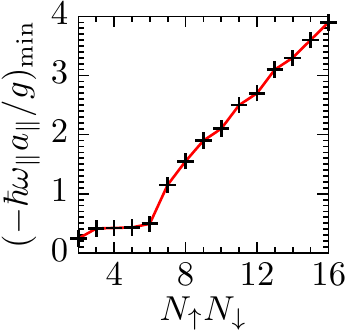}
 \end{tabular}
 \caption{(Color online)
   (a) The energy bands,
   following \figref{fig:TwoParticle}
   conventions.
   (b) The map of crossing events for the $n_{\text{band}}$-th excited
   molecular state with $N_{\uparrow}N_{\downarrow}$.  Blue denotes a single
   crossing, and red a double crossing.
   (c) The minimum interaction strength 
   to avoid the band crossing region.}
 \label{fig:MapPlot}
\end{figure}

The search for itinerant ferromagnetism in a cold atom gas has been plagued
by a competing loss process~\cite{Haller10,Pekker11,Sanner12}.  Several
models for loss have been put forward including two and three-body
models~\cite{Petrov03,Sala12,Pekker11,Sanner12}, and losses to states
excited with transverse COM
motion~\cite{Haller10,Sala13,Bolda05,Melezhik09}.  To conserve energy, both mechanisms require
the open channel to cross molecular bands. Our
Hamiltonian only displays avoided crossings between states with the same COM
quantum number. However, states with different COM motion could have avoided
crossings due to unforeseen perturbations such as an anharmonic
potential~\cite{Haller10,Sala13,Bolda05,Melezhik09}. To guarantee a
loss-free experiment, we fence off the region in which the open channel is
crossed or anti-crossed by any other state. This pessimistic approach
is robust to unforeseen perturbations that may alter the crossings but will
not significantly alter the positions of the bands.  We first focus on the
three-body system where we use \figref{fig:TwoParticle}(b) to define a loss
region as where the desired open channel $\ket{1/2,2,1}$ crosses the
molecular bound states.

The ground molecular bound state labeled (i) is lower than the entire open
channel $\ket{1/2,2,1}$ so its formation is prohibited by energy
conservation. The molecular bound state can be excited with COM motion,
giving rise to increasingly populous families of curves. The curves (ii) are
the first set of molecular bands to cross the state
$\ket{1/2,2,1}$. Further crossings from more excited
molecular bands occur up to the confinement induced resonance, prohibiting
experimentalists from looking for magnetic correlations within
$0\lesssim-\hbar\omega_{\parallel}a_{\parallel}/g\lesssim0.24$.  This region
contains the molecular bound state with COM motion excited into the second
transverse mode that is a significant cause of loss in an anharmonic
potential~\cite{Sala13,Sala12,Bolda05,Melezhik09}. Though the
definition of $g$ used to characterize the interaction strength does not
conform to the correct effective pseudopotential for the excited transverse
states~\cite{Bergeman03}, it properly describes the experimentally relevant
ground transverse states.

We note that it is possible to adiabatically transit across the region of
band crossing. Investigators can perform experiments on the $\ket{1/2,2,1}$
state either side of the shaded region in \figref{fig:ExpSetup}, but not
within it. A similar analysis of a system with four atoms reveals that
losses would block the region
$0\lesssim-\hbar\omega_{\parallel}a_{\parallel}/g\lesssim0.36$, and with
five atoms the range
$0\lesssim-\hbar\omega_{\parallel}a_{\parallel}/g\lesssim0.42$.

Exact diagonalization cannot accurately address larger systems. We therefore
turn to QMC for the open channel and the variational QMC for the molecular
band. In the Super-Tonks regime the energy difference between unpolarized
and polarized states is $\hbar\omega_{\parallel}N_{\uparrow}N_{\downarrow}$,
so we categorize states by $N_{\uparrow}N_{\downarrow}$. We focus on the
state that bounds the loss region, with the molecule having no COM motion,
and other atoms in higher energy orbitals compatible with the correct
non-interacting energy. With $N_{\uparrow}N_{\downarrow}=2$ in
\figref{fig:TwoParticle}(b) the excited molecule bands cross the upper
branch only once, whereas with $N_{\uparrow}N_{\downarrow}=8$ in
\figref{fig:MapPlot}(a) a molecular band, highlighted in green, crosses the
open channel twice. In \figref{fig:MapPlot}(b) we show whether the
$n_{\text{band}}$-th family of excited molecular bands crosses the open
channel once or twice. The double crossings first emerge at
$N_{\uparrow}N_{\downarrow}=7$ and become ubiquitous as
$N_{\uparrow}N_{\downarrow}$ rises. This leads to a proliferation in the
total number of band crossings, and as shown in \figref{fig:MapPlot}(c) a
dramatic rise in the minimum interaction strength $-\hbar
a_{\parallel}\omega_{\parallel}/g$ required to avoid band crossings.

\section{Discussion}

A quasi one-dimensional system containing a few fermionic atoms poses an
opportunity to explore ferromagnetic correlations. Discretization of the
energy levels offers the stabilization of a
ferromagnetic state without losses.  We have calculated the energy structure
and studied the 
ejection probabilities. 
Both the polaron energy and pair correlation function tend to the itinerant
limit when $N_{\text{tot}}\ge5$, whereas molecule losses restrict the
 observation of magnetic correlations to $N_{\text{tot}}\le6$.
Therefore, systems with $N_{\text{tot}}\in\{5,6\}$ could present an
opportunity to observe magnetic correlations driven by the Stoner mechanism.

\acknowledgments {The authors thank Gerhard Z\"urn, Thomas Lompe, Selim
  Jochim \& Stefan Baur for useful discussions. POB acknowledges the
  financial support of the EPSRC, and GJC the support of Gonville \& Caius
  College.}

\end{document}